\documentclass{PoS}

  \usepackage{amssymb}
  \usepackage{amsmath}
    \usepackage{amsfonts}
     \usepackage{epsfig}
      \usepackage{bm}

\usepackage[section]{placeins}

\usepackage{multirow}
\usepackage{ctable}
\usepackage{booktabs}
\usepackage{array}
\usepackage{tabularx}
\usepackage{xcolor}
\usepackage{pstricks}

\newcommand{\nn}{\nonumber}

\newcommand{\beq}{\begin{equation}}
\newcommand{\eeq}{\end{equation}}
\newcommand{\bea}{\begin{eqnarray}}
\newcommand{\eea}{\end{eqnarray}}
\newcommand{\beqa}{\begin{eqnarray}}
\newcommand{\eeqa}{\end{eqnarray}}

\definecolor{red}{rgb}{1,0,0}

\def\be{\begin{equation}}
\def\ee{\end{equation}}
\numberwithin{equation}{section}

\title{Searching for optimal conditions for exploration of double-parton scattering in four-jet production\\ at the LHC}

\ShortTitle{Double-parton scattering in four-jet production}

\author{\speaker{Rafa{\l} Maciu{\l}a}\thanks{The work has been supported by the Polish National Science Center grant
DEC-2014/15/B/ST2/02528.}\\
        Institute of Nuclear
Physics, Polish Academy of Sciences, Radzikowskiego 152,\\ PL-31-342 Krak{\'o}w, Poland\\
        E-mail: \email{rafal.maciula@ifj.edu.pl}}

\author{Antoni Szczurek\\
Institute of Nuclear
Physics, Polish Academy of Sciences, Radzikowskiego 152,\\ PL-31-342 Krak{\'o}w, Poland\\
        E-mail: \email{antoni.szczurek@ifj.edu.pl}}

\abstract{We discuss the double-parton scattering (DPS) effects in four-jet production at the LHC. A special attention is given to 
the optimization of kinematical conditions in order to enhance the relative contribution of DPS in four-jet sample. The calculations
of single-parton and double-parton scattering components are done in the high-energy (or $k_{T}$)-factorization approach. Here we follow our recent developments of relevant methods and tools that are implemented in the AVHLIB Monte Carlo library.   
Several differential distributions are calculated and carefully discussed in the context of recent and future searches for DPS
effects at the LHC, for both $\sqrt{s} = 7$ and $13$ TeV energies. The dependences of the relative DPS amount is studied as a function of rapidity of jets, rapidity distance, and various azimuthal correlations between jets. The regions with an enhanced DPS contribution are identified.}

\FullConference{XXIV International Workshop on Deep-Inelastic Scattering and Related Subjects\\
		11-15 April, 2016\\
		DESY Hamburg, Germany}

\begin{document}

\section{Introduction}

Four-jet production seems a natural case to look for hard double-parton scattering (DPS) effects (see e.g. Ref.~\cite{Kutak:2016ukc} and references therein). A year ago we have analyzed how to find optimal
conditions for the observation and exploration of DPS 
effects in four-jet production \cite{Maciula:2015vza}. In this analysis only 
the leading-order (LO) collinear approach was applied both to single and double-parton scattering mechanisms.

Very recently, we have performed for the first time a calculation of 
four-jet production for both single-parton and double-parton mechanism 
within $k_T$-factorization \cite{Kutak:2016mik}. 
It was shown that the effective inclusion of higher-order effects leads 
to a substantial damping of the double-scattering contribution 
with respect to the SPS one, especially for symmetric (identical) cuts on the transverse momenta of all jets. 

So far, most practical calculations of DPS contributions were performed 
within the so-called factorized ansatz. 
In this approach, the cross section for DPS is a product 
of the corresponding cross sections for single-parton scatterings (SPS). 
This is a phenomenologically motivated approximation which is not well under control yet. 
A better formalism exists in principle, but predictions are not easy, 
as they require unknown input(s), \textit{e.g.} double-parton distributions that should contain informations about space-configuration, spin, colour or flavour correlations between the two partons in one hadron \cite{Diehl:2011yj}. These objects are explored to a far lesser extent than the standard single PDFs. 
However, the factorized model seems to be a reasonable tool to collect
a sufficient amount of empirical facts to draw practical conclusions about possible identification of the DPS effects
in several processes. 

As discussed in Ref.~\cite{Maciula:2015vza}, jets with low cuts on the transverse momenta and a large rapidity separation 
seem more promising than others in exploring DPS effects in four-jet production.
In the following we shall concentrate on the study of this and other more optimal observables
to pin down DPS contributions in the $k_T$-factorization framework.

\section{A sketch of the theoretical formalism}

The theoretical formalism used to obtain the following predictions has been discussed in detail in our previous paper \cite{Kutak:2016mik}.
All details related to the scattering amplitudes with off-shell initial state partons as well as with the Transverse Momentum Dependent parton distribution functions (TMDs) can be found there.

Here we only very briefly recall the high-energy (or $k_{T}$)-factorization (HEF) formula for the calculation of the inclusive partonic four-jet cross section at the Born level: 
\bea
\sigma^B_{4-jets} 
&=& 
\sum_{i,j} \int \frac{dx_1}{x_1}\,\frac{dx_2}{x_2}\, d^2 k_{T1} d^2 k_{T2}\,  \mathcal{F}_i(x_1,k_{T1},\mu_F)\, \mathcal{F}_j(x_2,k_{T2},\mu_F) \nn \\
&&
\hspace{-25mm}
\times \frac{1}{2 \hat{s}} \prod_{l=i}^4 \frac{d^3 k_l}{(2\pi)^3 2 E_l} \Theta_{4-jet} \, (2\pi)^4\, \delta\left( x_1P_1 + x_2P_2 + \vec{k}_{T\,1}+ \vec{k}_{T\,2} - \sum_{l=1}^4 k_i \right)\, 
\overline{ \left| \mathcal{M}(i^*,j^* \rightarrow 4\, \text{part.})
\right|^2 } \, . \nn \\
\label{kt_cross}
\eea
Here $\mathcal{F}_i(x_k,k_{Tk},\mu_F)$ is the TMD for a given parton type, $x_k$ are the longitudinal
momentum fractions, $\mu_F$ is a factorization scale, $\vec{k}_{Tk}$ the parton's transverse momenta.
$\mathcal{M}(i^*,j^* \rightarrow 4\, \text{part.})$ is the gauge invariant matrix element for $2\rightarrow 4$ particle scattering with two initial off-shell partons. They are evaluated numerically with the AVHLIB~\cite{Bury:2015dla} Monte Carlo library.
In the calculation, the scales are set to
$\mu_F=\mu_R= \frac{\hat{H}_T}{2} = \frac{1}{2} \sum_{l=1}^4 k_T^l$\footnote{We use the $\hat{H}_T$ notation to refer to the energies of the final state partons, not jets, despite this is obviously the same thing in a LO analysis.}.

The so called pocket formula for DPS cross sections (for a four-parton final state) is given by
\beq
\frac{d \sigma^{B}_{4-jet,DPS}}{d \xi_1 d \xi_2} = 
\frac{m}{\sigma_{eff}} \sum_{i_1,j_1,k_1,l_1;i_2,j_2,k_2,l_2} 
\frac{d \sigma^B(i_1 j_1 \rightarrow k_1 l_1)}{d \xi_1}\, \frac{d \sigma^B(i_2 j_2 \rightarrow k_2 l_2)}{d \xi_2} \, ,
\eeq
where the $\sigma(a b \rightarrow c d)$ cross sections are obtained by
restricting (\ref{kt_cross}) to a single channel and the symmetry factor $m$ is $1/2$ if the two hard
scatterings are identical, to avoid double counting.
Finally, $\xi_1$ and $\xi_2$ stand for generic kinematical variables for the first and second scattering, respectively.
The effective cross section $\sigma_{eff}$ can be interpreted as a measure of the transverse correlation of the two partons inside 
the hadrons, whereas the possible longitudinal correlations are usually neglected.
In the numerical calculations here we use $\sigma_{eff}$ = 15 mb that is the typical value known from the world systematics \cite{Proceedings:2016tff}. 
\section{Numerical results}

\begin{figure}[!h]
\begin{minipage}{0.47\textwidth}
 \centerline{\includegraphics[width=1.0\textwidth]{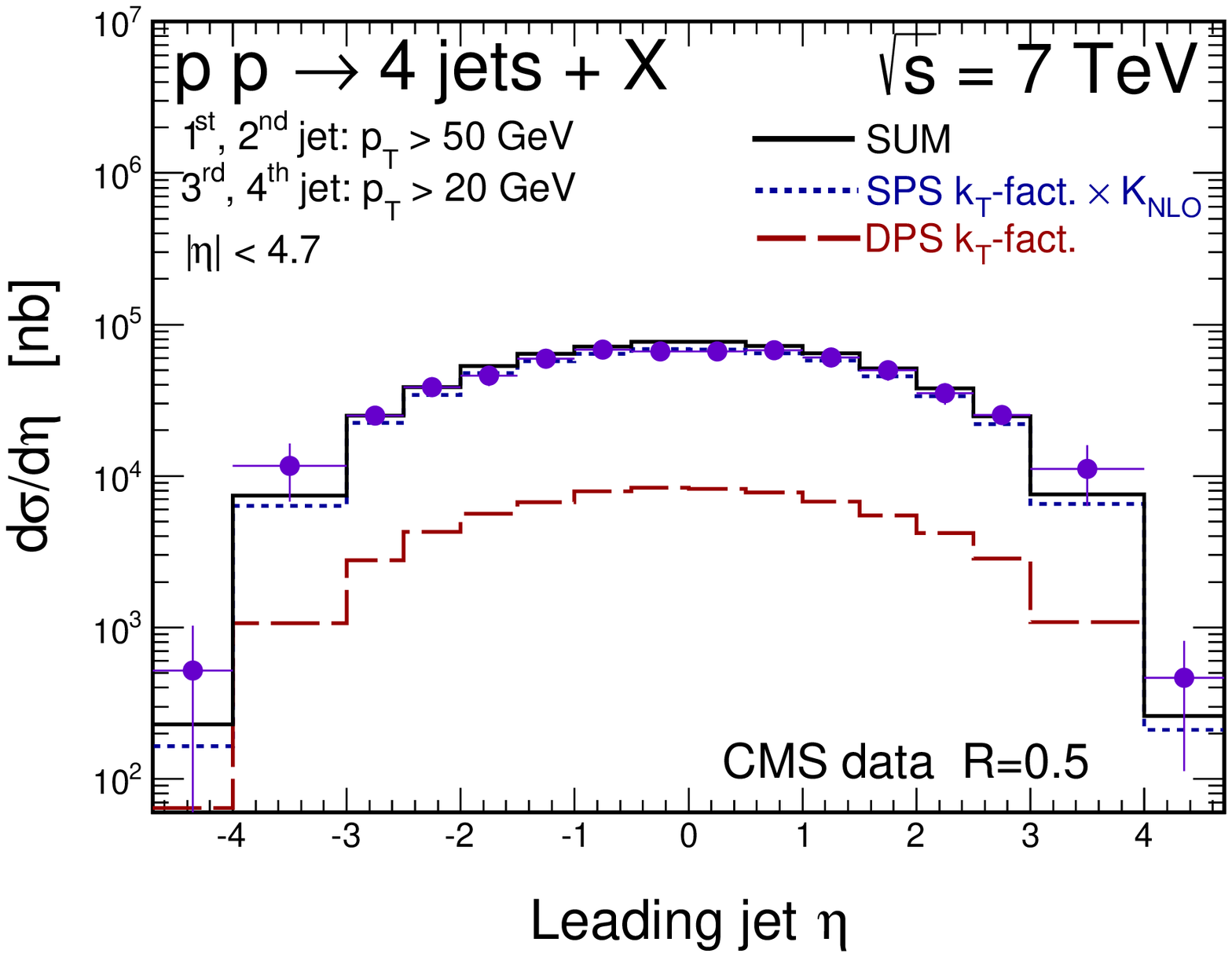}}
\end{minipage}
\hspace{0.5cm}
\begin{minipage}{0.47\textwidth}
 \centerline{\includegraphics[width=1.0\textwidth]{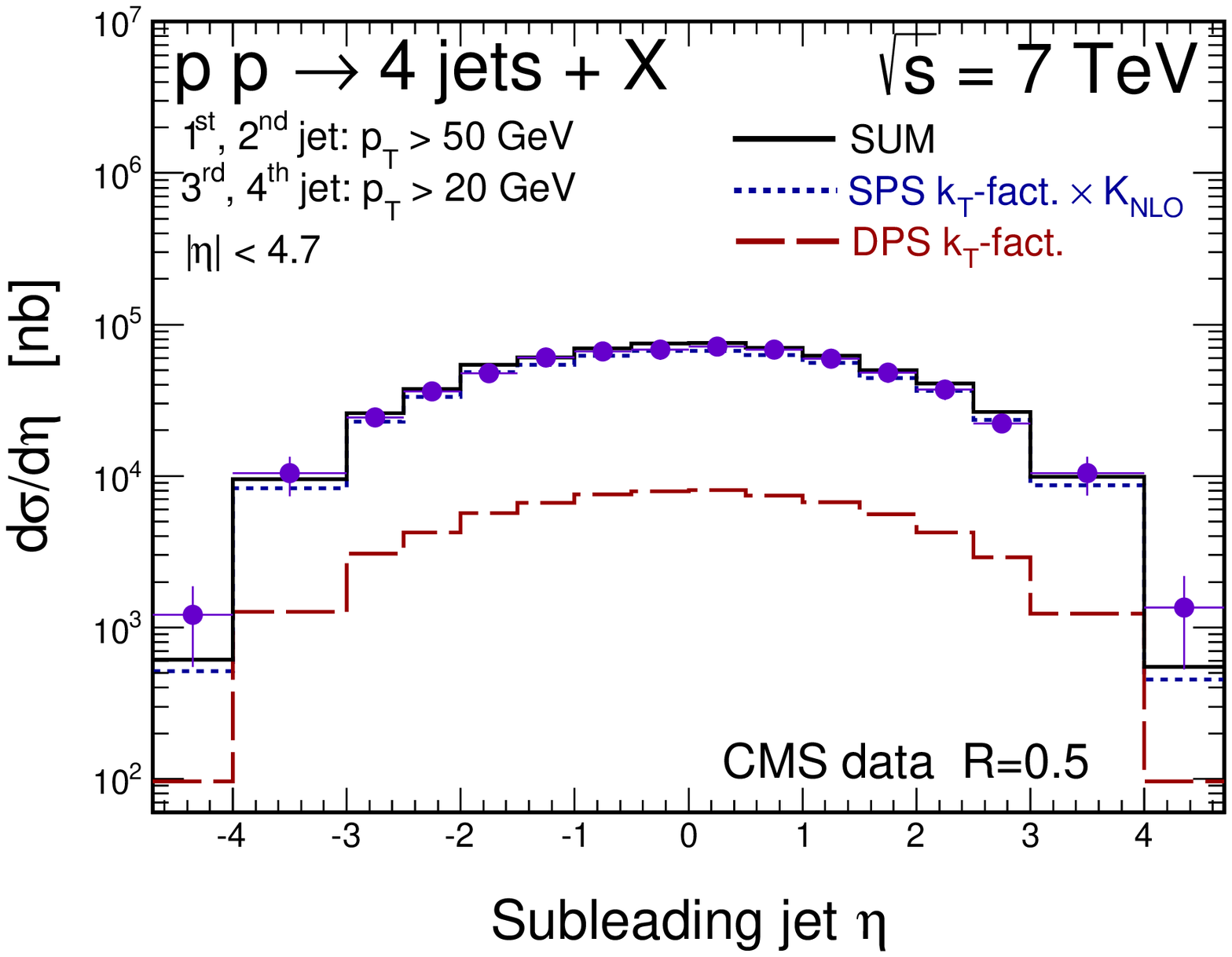}}
\end{minipage}
   \caption{
\small Rapidity distribution of the leading and subleading jets.
The SPS contribution is shown by the dotted line while the DPS contribution by the dashed line.
 }
 \label{fig:CMS_y_distributions}
\end{figure}

We start with a comparison of our numerical predictions with existing experimental data for relatively low cuts on jet transverse momenta.
In this context, the CMS experimental multi-jet analysis \cite{Chatrchyan:2013qza} is the most relevant as it uses sufficiently soft cuts on the jet transverse momenta for DPS. The cuts are in this case $|p_T| > 50$ GeV for the two hardest jets 
and $|p_T| > 20$ GeV for the third and fourth ones; the rapidity region is defined by $|\eta| < 4.7$ 
and the constraint on the jet cone radius parameter is $\Delta R >0.5$.
The overall situation is shown in Fig.~\ref{fig:CMS_y_distributions}, 
where we plot rapidity distributions for leading and subleading jets ordered by their $p_{T}$'s.

The $k_T$-factorization approach includes higher-order corrections 
through the resummation in the TMDs even at the Born level. However, within this framework fixed-order loop effects are not taken into account.
Therefore, we allow for a $K$-factor for the calculation of the SPS component.
The NLO $K$-factors are known to be smaller than unity
for 3- and 4-jet production in the collinear approximation case \cite{Bern:2011ep}. 
To describe the CMS data, we also need $K$-factors smaller than unity for the SPS contributions, as expected.
In contrast to the 4-jet case, the NLO predictions for the 2-jet inclusive cross section 
are further away from the measured value than the LO ones \cite{Bern:2011ep}. The 2-jet $K$-factor is known to be about $1.2$, however it enters squared in the case of the DPS calculations. Therefore, in order to ensure that the predicted DPS effects are not overestimated, we ignore $K$-factors for the DPS.  

\begin{figure}[!h]
\begin{minipage}{0.47\textwidth}
 \centerline{\includegraphics[width=1.0\textwidth]{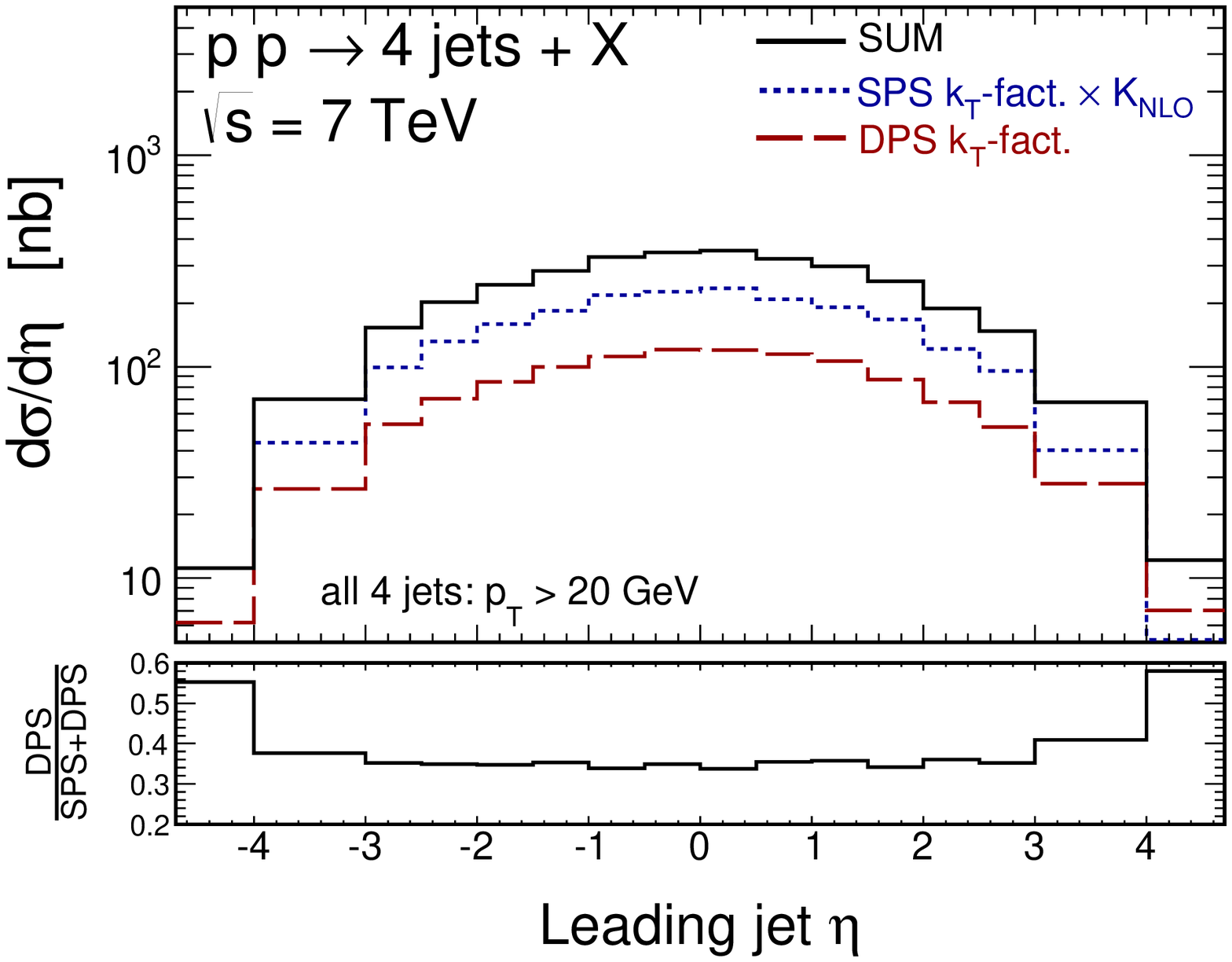}}
\end{minipage}
\hspace{0.5cm}
\begin{minipage}{0.47\textwidth}
 \centerline{\includegraphics[width=1.0\textwidth]{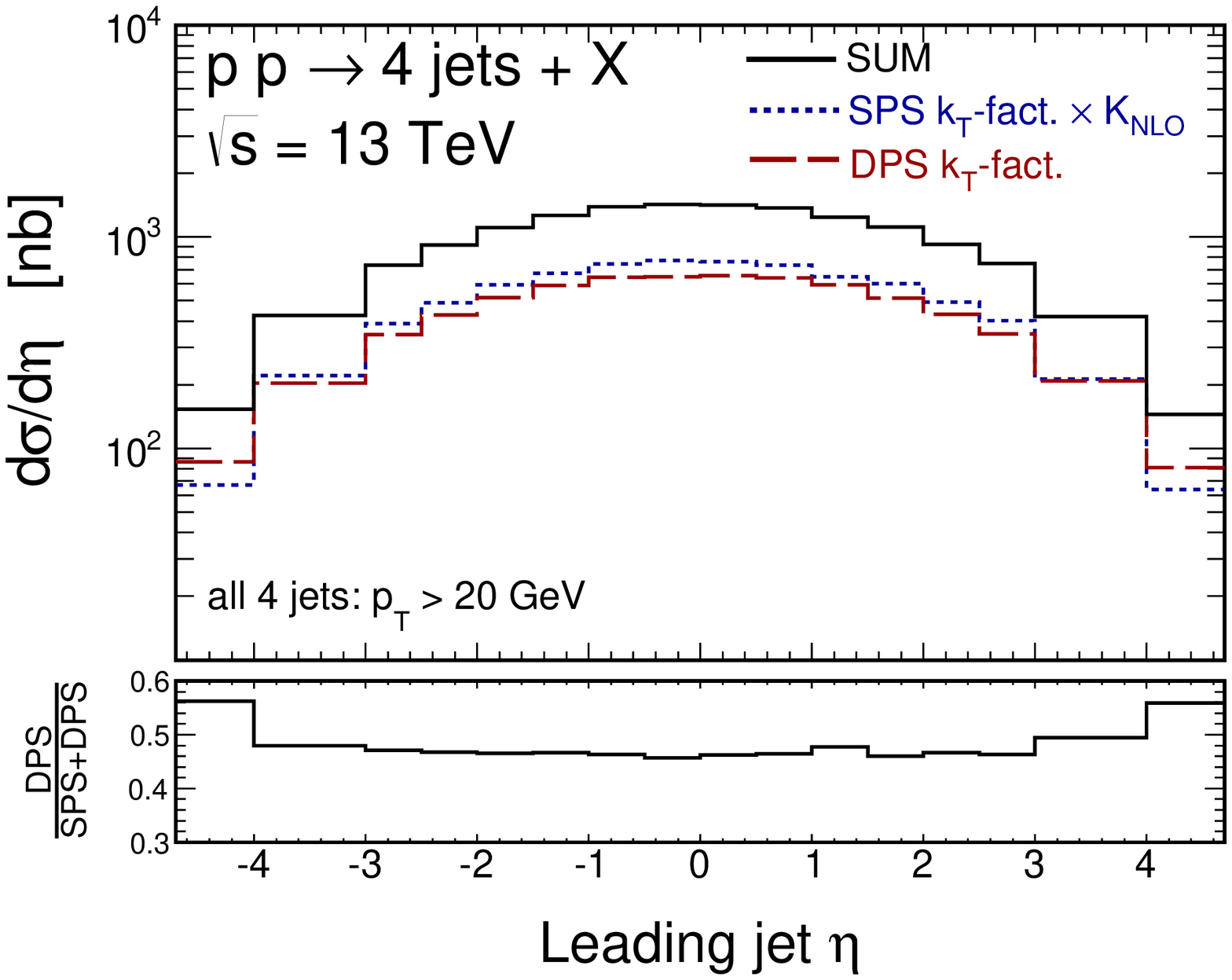}}
\end{minipage}
   \caption{
\small Rapidity distribution of leading jet for $\sqrt{s}$ = 7 TeV
(left column) and $\sqrt{s}$ = 13 TeV (right column) for
the symmetric cuts. The SPS contribution
is shown by the dotted line while the DPS contribution by the dashed line.
The relative contribution of DPS is shown in the extra lower panels.
 }
 \label{fig:dsig_dy_symmetric_20GeV}
\end{figure}

In the following we introduce a set of observables that we find particularly convenient to
identify DPS effects in four-jet production.
Here we use completely symmetric cuts, $p_T > 20$ GeV, for all the four leading jets.
The cuts on rapidity and jet radius parameter stay the same as for the CMS case.
In Fig.~\ref{fig:dsig_dy_symmetric_20GeV} we show our predictions
for the rapidity distributions. In contrast to the previous case (Fig.~\ref{fig:CMS_y_distributions}), where harder cuts on the two hardest jets were used, the shapes of the SPS and DPS rapidity distributions 
are rather similar. There is only a small relative enhancement of the DPS contribution for larger jet rapidities.

As it was shown first in Ref.~\cite{Maciula:2014pla} in the context of Mueller-Navelet jet production, and then
repeated in Ref.~\cite{Maciula:2015vza} for four-jet studies in the LO collinear approach, there are two potentially
useful observables for DPS effects, namely the maximum rapidity distance 
\begin{equation}
\Delta \text{Y} \equiv max_{\substack{i,j \in\{1,2,3,4\}\\i \neq j }} |\eta_i-\eta_j |
\end{equation}
and the azimuthal correlations between the jets which are most remote in rapidity
\begin{equation}
\varphi_{jj} \equiv | \varphi_i -\varphi_j |  \, , \quad \text{for}  \quad |\eta_i - \eta_j | = \Delta \text{Y} \, .
\end{equation}

\begin{figure}[!h]
\begin{minipage}{0.47\textwidth}
 \centerline{\includegraphics[width=1.0\textwidth]{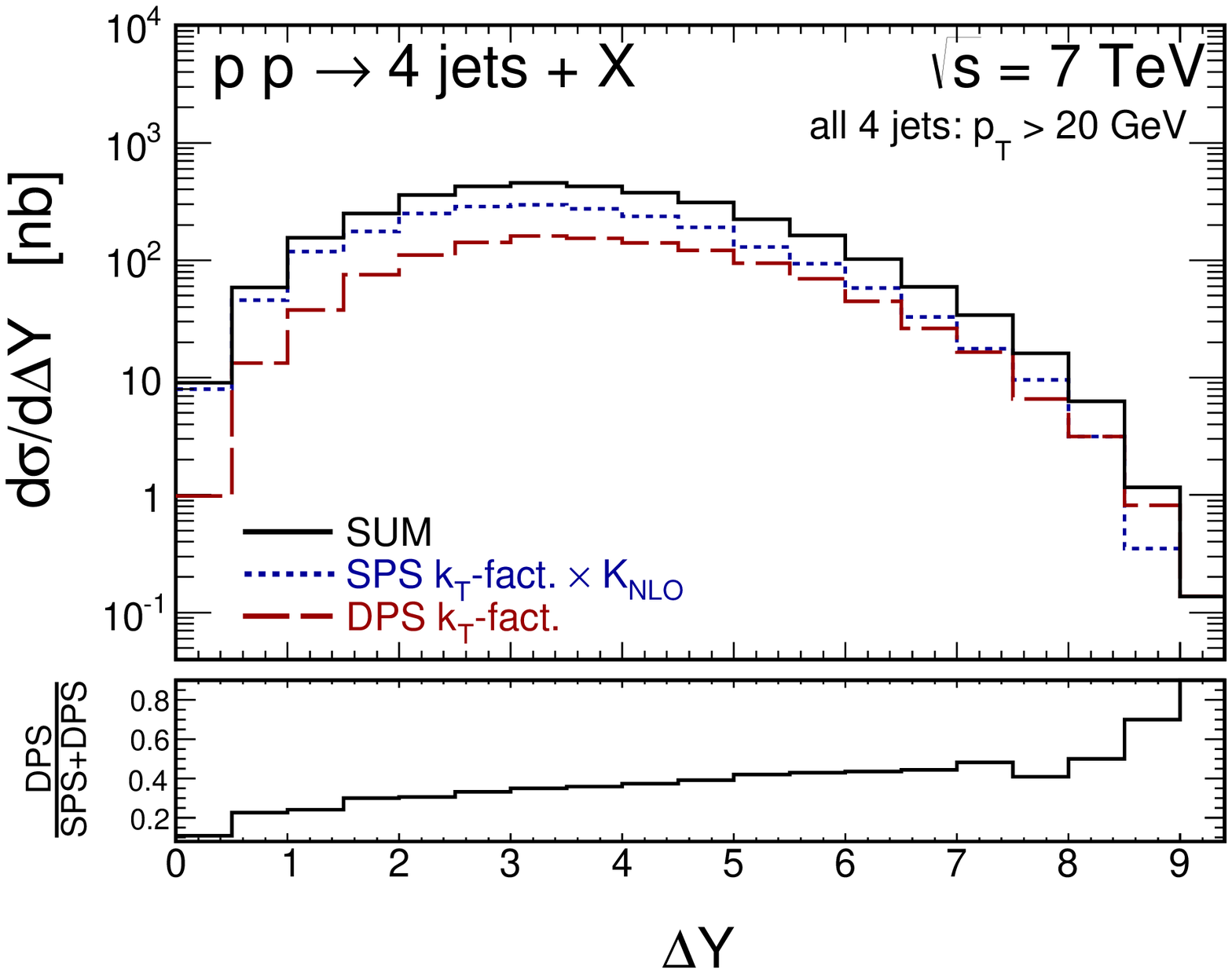}}
\end{minipage}
\hspace{0.5cm}
\begin{minipage}{0.47\textwidth}
 \centerline{\includegraphics[width=1.0\textwidth]{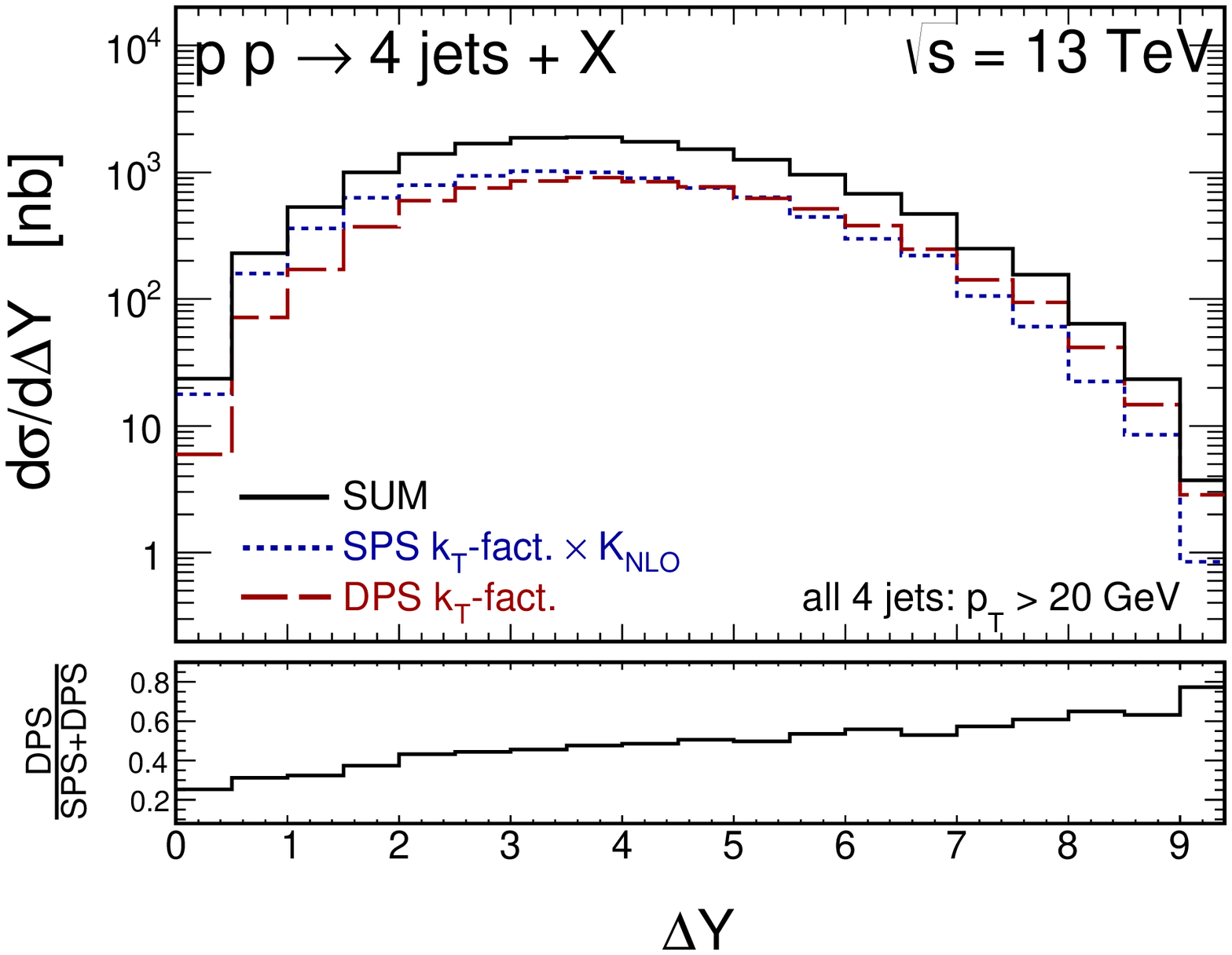}}
\end{minipage}
   \caption{
\small Distribution in rapidity distance between the most remote jets for 
the symmetric cut with $p_T >$ 20 GeV 
for $\sqrt{s}$ = 7 TeV (left) and $\sqrt{s}$ = 13 TeV (right).
The SPS contribution is shown by the dotted line while 
the DPS contribution by the dashed line.
The relative contribution of DPS is shown in the extra lower panels.
 }
 \label{fig:dsig_dydiff}
\end{figure}

One can see in Fig.~\ref{fig:dsig_dydiff} that the relative DPS contribution increases 
with $\Delta \text{Y}$ which, for the CMS collaboration, can be as large as 9.4.
At $\sqrt{s}$ = 13 TeV the DPS component dominates over the SPS contribution for $\Delta \text{Y} > 6$.
A potential failure of the SPS contribution to describe such a plot in this region
would be a signal of the presence of a sizable DPS contribution.

\begin{figure}[!h]
\begin{minipage}{0.47\textwidth}
 \centerline{\includegraphics[width=1.0\textwidth]{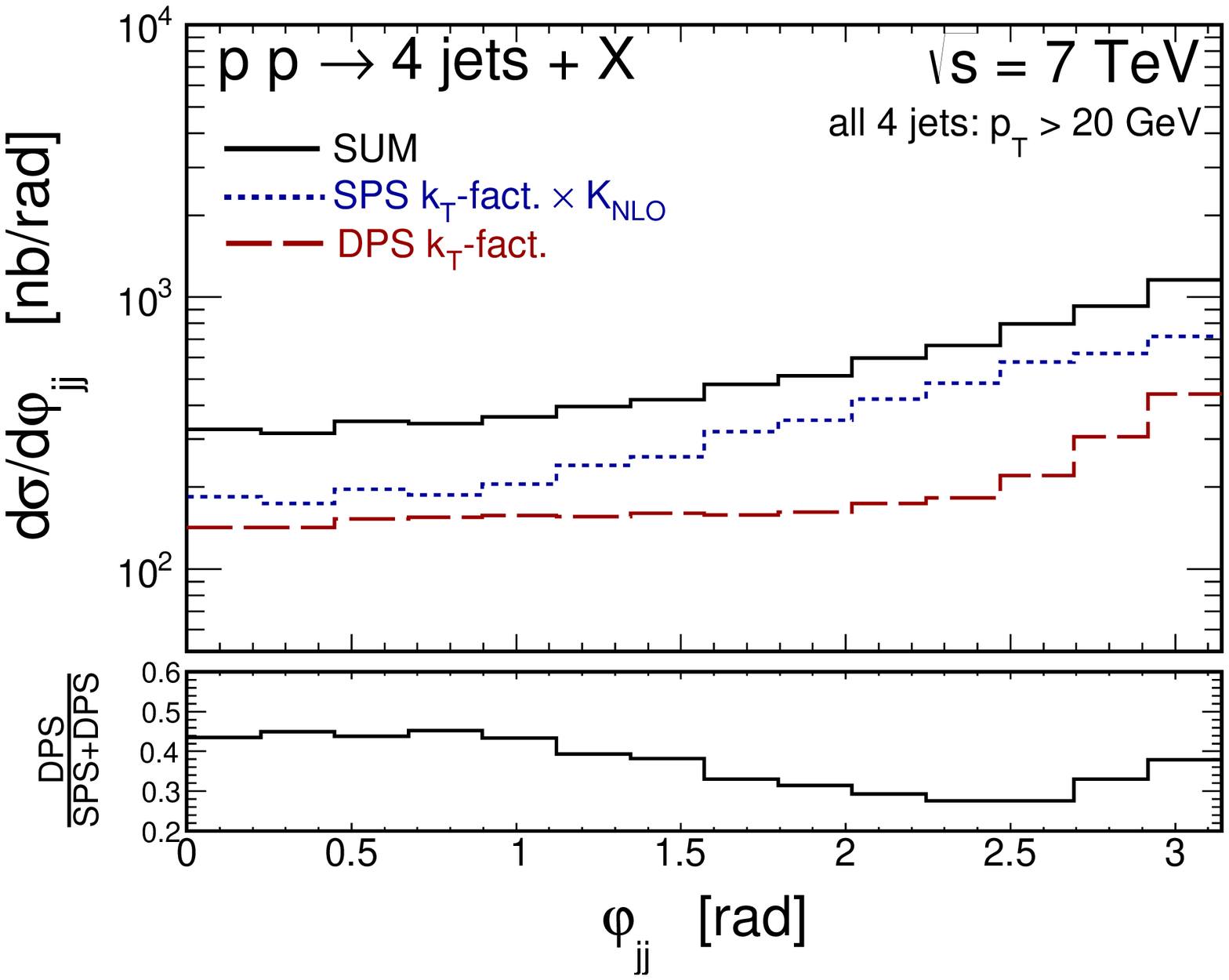}}
\end{minipage}
\hspace{0.5cm}
\begin{minipage}{0.47\textwidth}
 \centerline{\includegraphics[width=1.0\textwidth]{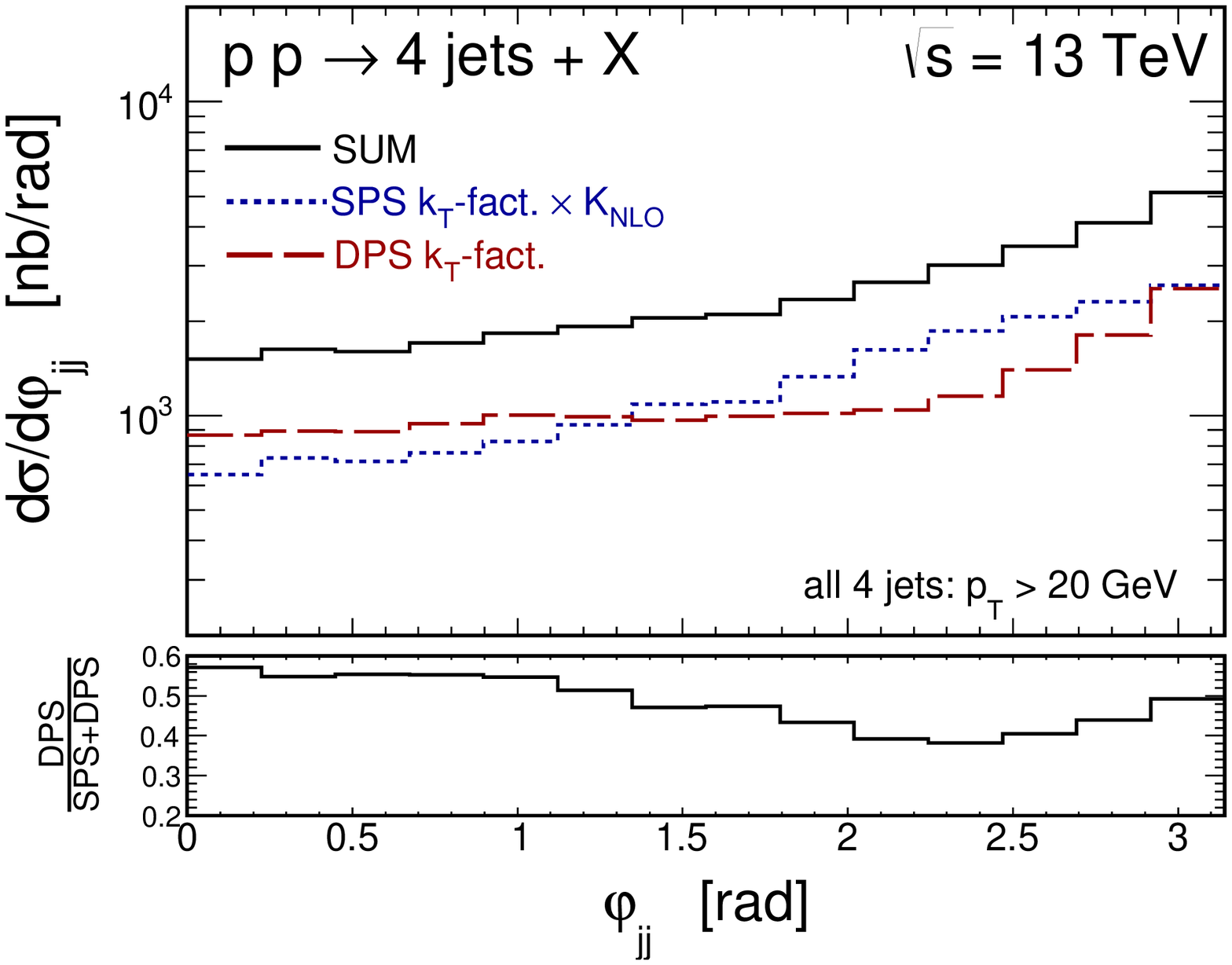}}
\end{minipage}
   \caption{
\small Distribution in relative azimuthal angle between the most remote jets for 
the symmetric cut with $p_T >$ 20 GeV 
for $\sqrt{s}$ = 7 TeV (left) and $\sqrt{s}$ = 13 TeV (right).
The SPS contribution is shown by the dotted line while 
the DPS contribution by the dashed line.
The relative contribution of DPS is shown in the extra lower panels.
 }
 \label{fig:dsig_dphijj}
\end{figure}

Figure~\ref{fig:dsig_dphijj} shows azimuthal correlations between the jets most remote in rapidity. 
While at $\sqrt{s}$~=~7~TeV the SPS contribution is always larger than the DPS one, 
at $\sqrt{s}$ = 13 TeV the DPS component dominates over the SPS contribution for $\varphi_{jj} < \pi/2$. 

\begin{figure}[!h]
\begin{minipage}{0.47\textwidth}
 \centerline{\includegraphics[width=1.0\textwidth]{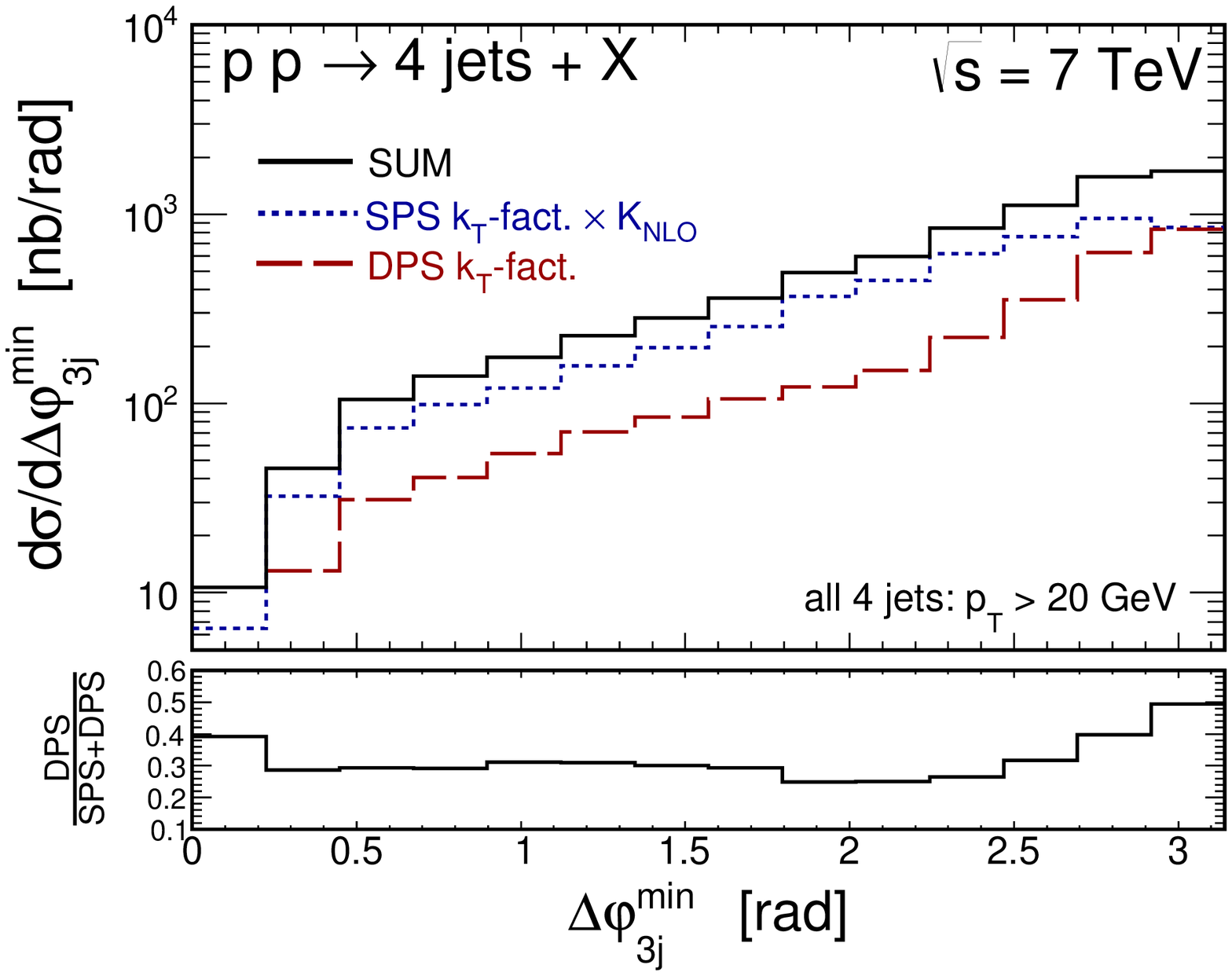}}
\end{minipage}
\hspace{0.5cm}
\begin{minipage}{0.47\textwidth}
 \centerline{\includegraphics[width=1.0\textwidth]{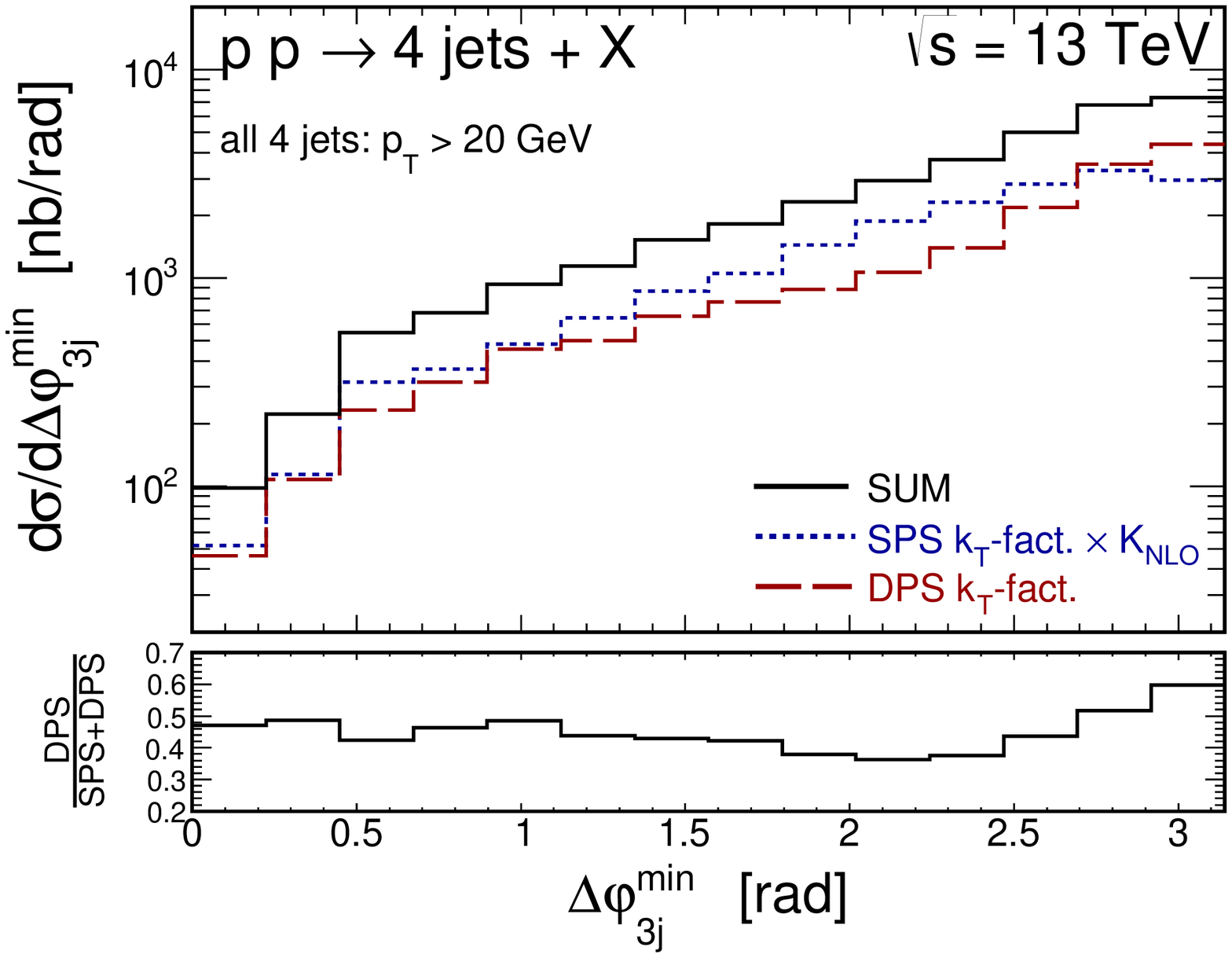}}
\end{minipage}
   \caption{
\small Distribution in $\Delta \varphi_{3j}^{min}$ angle for 
the symmetric cut with $p_T >$ 20 GeV 
for $\sqrt{s}$ = 7 TeV (left) and $\sqrt{s}$ = 13 TeV (right).
The SPS contribution is shown by the dotted line while 
the DPS contribution by the dashed line.
The relative contribution of DPS is shown in the extra lower panels.
 }
 \label{fig:dsig_dphi_3j}
\end{figure}

We also find that another variable, introduced in the high transverse momenta analysis 
of four jets production presented in Ref.~\cite{Aad:2015nda}, can be very interesting for 
the examination of DPS effects:
\begin{equation}
\Delta \varphi_{3j}^{min} \equiv min_{\substack{i,j,k \in\{1,2,3,4\}\\i 
\neq j \neq k}}\left(|\varphi_i - \varphi_j|+| \varphi_j - \varphi_k|\right) \, .
\label{DeltaPhiMin}
\end{equation}
As three out of four azimuthal angles are always entering in (\ref{DeltaPhiMin}), configurations
featuring one jet recoiling against the other three are necessarily characterised by lower
values of $\Delta \varphi_{3j}^{min}$ with respect to the two-against-two topology;
the minimum, in fact, will be obtained in the first case for $i,j,k$ 
denoting the three jets in the same half hemisphere, 
whereas no such thing is possible for the second configuration.
Obviously, the first case would be allowed only by SPS in a collinear tree-level framework,
whereas the second would be enhanced by DPS. 
In the $k_T$-factorization approach, 
this situation is smeared out by the presence of transverse momenta of the initial state partons. 
For our TMDs, the corresponding distributions are shown in Fig.~\ref{fig:dsig_dphi_3j}.
We do not see such obvious effects in the case of $k_{T}$-factorization.

\section{Conclusions}

In this presentation we have discussed how to look at the DPS effects and how to maximize their role in four jet production. 
We found that, for sufficiently small cuts on the transverse momenta, DPS effects are enhanced relative to the SPS contribution:
when rapidities of jets  are large, for large rapidity distances between the most remote jets, for small azimuthal angles between the two jets most remote in rapidity and/or for large values of the $\Delta\varphi_{3j}^{min}$ variable. For more details we refer the interested reader to our regular article \cite{Kutak:2016ukc}.


\begin{thebibliography}{99}
\bibitem{Kutak:2016ukc} 
  K.~Kutak, R.~Maciula, M.~Serino, A.~Szczurek and A.~van Hameren,
  arXiv:1605.08240 [hep-ph].
  
\bibitem{Maciula:2015vza} 
  R.~Maciula and A.~Szczurek,
  Phys.\ Lett.\ B {\bf 749}, 57 (2015)
  [arXiv:1503.08022 [hep-ph]].
  
\bibitem{Kutak:2016mik} 
  K.~Kutak, R.~Maciula, M.~Serino, A.~Szczurek and A.~van Hameren,
  JHEP {\bf 1604}, 175 (2016).

\bibitem{Diehl:2011yj} 
  M.~Diehl, D.~Ostermeier and A.~Schafer,
  JHEP {\bf 1203}, 089 (2012)
  [arXiv:1111.0910 [hep-ph]].
  
\bibitem{Bury:2015dla} 
  M.~Bury and A.~van Hameren,
  Comput.\ Phys.\ Commun.\  {\bf 196}, 592 (2015).

\bibitem{Proceedings:2016tff} 
  H.~Jung, D.~Treleani, M.~Strikman and N.~van Buuren,
  DESY-PROC-2016-01.
  
\bibitem{Chatrchyan:2013qza} 
  S.~Chatrchyan {\it et al.} [CMS Collaboration],
  Phys.\ Rev.\ D {\bf 89}, 092010 (2014).

\bibitem{Bern:2011ep} 
  Z.~Bern {\it et al.},
  Phys.\ Rev.\ Lett.\  {\bf 109}, 042001 (2012)
  [arXiv:1112.3940 [hep-ph]].


\bibitem{Maciula:2014pla} 
  R.~Maciula and A.~Szczurek,
  Phys.\ Rev.\ D {\bf 90}, 014022 (2014)
  [arXiv:1403.2595 [hep-ph]].
  
\bibitem{Aad:2015nda} 
  G.~Aad {\it et al.} [ATLAS Collaboration],
  JHEP {\bf 1512}, 105 (2015)
  [arXiv:1509.07335 [hep-ex]].



\end{thebibliography}
\end{document}